\documentstyle[stwol]{article}

\input{psfig}
\def\Journal#1#2#3#4{{#1} {\bf #2}, #3 (#4)}

\def\NPB{{\em Nucl.~Phys.}~B}
\def\PLB{{\em Phys.~Lett.}~B}
\def\PRL{\em Phys.~Rev.~Lett.}
\def\PRD{{\em Phys.~Rev.}~D}

\def\ra{\rightarrow}

\def\be{\begin{equation}}
\def\ee{\end{equation}}
\def\bea{\begin{eqnarray}}
\def\eea{\end{eqnarray}}
\def\ba{\begin{array}}
\def\ea{\end{array}}
\def\simge{\mathrel{%
   \rlap{\raise 0.511ex \hbox{$>$}}{\lower 0.511ex \hbox{$\sim$}}}}
\def\simle{\mathrel{
   \rlap{\raise 0.511ex \hbox{$<$}}{\lower 0.511ex \hbox{$\sim$}}}}
 
\def\slashchar#1{\setbox0=\hbox{$#1$}           % set a box for #1
   \dimen0=\wd0                                 % and get its size
   \setbox1=\hbox{/} \dimen1=\wd1               % get size of /
   \ifdim\dimen0>\dimen1                        % #1 is bigger
      \rlap{\hbox to \dimen0{\hfil/\hfil}}      % so center / in box
      #1                                        % and print #1
   \else                                        % / is bigger
      \rlap{\hbox to \dimen1{\hfil$#1$\hfil}}   % so center #1
      /                                         % and print /
   \fi}                                         %
\def\ts{\thinspace}

\def\ol{\bar}

\def\CL{{\cal L}}

\def\CO{{\cal O}}

\def\atc{\alpha_{TC}}
\def\aqcd{\alpha_{S}}
\def\atro{\alpha_{\rho_T}}
\def\Few{F_\pi}
\def\Ntc{N_{TC}}
\def\suc{SU(3)}

\def\sutc{SU(\Ntc)}

\def\getc{g_{ETC}}

\def\kslash{\raise.15ex\hbox{/}\kern-.57em k}
\def\LTC{\Lambda_{TC}}

\def\METC{M_{ETC}}

\def\condtc{\langle \ol T T \rangle_{TC}}
\def\condetc{\langle \ol T T \rangle_{ETC}}
\def\tro{\rho_{T1}}
 
\def\troct{\rho_{T8}} 
\def\tropm{\rho_{T1}^\pm}

\def\troz{\rho_{T1}^0}
\def\tpi{\pi_T}
\def\tpipm{\pi_T^\pm}
\def\tpimp{\pi_T^\mp}
\def\tpip{\pi_T^+}
\def\tpim{\pi_T^-}
\def\tpiz{\pi_T^0}
\def\tpipr{\pi_T^{0 \prime}}
\def\etat{\eta_T}
\def\octpi{\pi_{T8}}
\def\octpipm{\pi_{T8}^\pm}

\def\octpiz{\pi_{T8}^0}
\def\toppi{\pi_t}
\def\toppip{\pi_t^+}
\def\toppim{\pi_t^-}
\def\toppipm{\pi_t^\pm}
\def\toppiz{\pi_t^0}

\def\Mv{M_{V_8}}
\def\Mzp{M_{Z'}}
\def\tpilq{\pi_{L \ol Q}}

\def\tpiql{\pi_{Q \ol L}}
\def\tpiun{\pi_{U \ol N}}
\def\tpiue{\pi_{U \ol E}}
\def\tpidn{\pi_{D \ol N}}
\def\tpide{\pi_{D \ol E}}

\def\jet{\rm jet}
\def\jets{\rm jets}

\def\mev{{\rm MeV}}
\def\gev{{\rm GeV}}
\def\tev{{\rm TeV}}

\def\pb{{\rm pb}}
\def\cm{{\rm cm}}
\def\ipb{{\rm pb}^{-1}}
\def\fb{{\rm fb}}
\def\half{{\textstyle{ { 1\over { 2 } }}}}
\def\third{{\textstyle{ { 1\over { 3 } }}}}

\def\eighth{{\textstyle{{1\over {8}}}}}
\def\twothirds{{\textstyle{ { 2\over { 3 } }}}}
\def\sixth{{\textstyle{ { 1\over { 6 } }}}}

\begin{document}

\title{NON-SUPERSYMMETRIC EXTENSIONS OF THE STANDARD MODEL}

\author{KENNETH LANE}

\address{Department of Physics, Boston University, 590 Commonwealth Ave,
Boston, MA 02215, USA}

\twocolumn[\maketitle\abstracts{The motivations for studying dynamical
scenarios of electroweak and flavor symmetry breaking are reviewed and
the latest ideas, especially topcolor-assisted technicolor, are
summarized. Several technicolor signatures at the Tevatron and Large
Hadron Collider are described and it is emphasized that all of them are
well within the reach of these colliders.}]

\section{Introduction}\label{sec:intro}

The title of my talk was chosen by the organizers and, while it was not
their intention, they have defined my subject by what it is not. That
leaves it for me to define what it is. So, in this talk
``non-supersymmetric extensions of the standard model'' means {\sl
Dynamical Electroweak and Flavor Symmetry Breaking.} To be specific, I will
discuss aspects of technicolor~\cite{tcref} and extended
technicolor~\cite{etcsd}$^{,\ts}$\cite{etceekl}.

I begin in Sec.~\ref{sec:motivate} by reiterating why it is still important
to study scenarios in which electroweak and flavor symmetry are broken by
strong dynamics at moderate, {\em accessible} energy scales. This is
followed in Sec.~\ref{sec:techni} by a review of technicolor and extended
technicolor, focusing on the more modern aspects---walking technicolor,
multiscale technicolor, and topcolor-assisted technicolor. In
Sec.~\ref{sec:signals}, I will discuss several important signatures of
these strong dynamics that can be sought over the next 10-15 years at the
upgraded Tevatron Collider and the Large Hadron Collider. For the most
part, these signatures involve the production of technihadrons $\rho_T$ and
$\omega_T$ and their decay into pairs of technipions, $\tpi\tpi$, $W_L
\tpi$ and $Z_L \tpi$, and possibly dijets. I restrict myself to these
hadron colliders not only because they are the only new high-energy
machines anywhere near the real axis,~\footnote{Some might view my saying
this as the kiss of death.} but also because they have the greatest reach
of all machines under consideration for the unknown physics of the
TeV~energy scale.

\section{Why Study Strong Electroweak and Flavor
Dynamics?}\label{sec:motivate}

The theoretical elements of the standard $\suc\otimes SU(2)\otimes U(1)$
gauge model of strong and electroweak interactions have been in place for
almost 25~years.~\cite{sm} In all this time, the standard model has
withstood extremely stringent experimental tests, the latest round being
described at this conference by Brock,~\cite{brock} Tipton,~\cite{tipton}
and Blondel.~\cite{blondel} Down to distances of at least $10^{-16}\,\cm$,
the basic constituents of matter are known to be spin-$\half$ quarks and
leptons. These interact via the exchange of spin-one gauge bosons: the
massless gluons of QCD and the massless photon and massive $W^\pm$ and
$Z^0$ bosons of electroweak interactions. There are six flavors each of
quarks and leptons---identical except for mass, charge and color---grouped
into three generations.

The fact that the QCD gauge symmetry is exact in both the Lagrangian
and the ground state of the theory implies that quarks and gluons are
confined at large distances into color-singlet hadrons and that they are
almost noninteracting at small distances. However, confinement and
asymptotic freedom are not the only dynamical outcomes for gauge theories.
Even though gauge bosons necessarily appear in the Lagrangian without mass,
interactions can make them heavy. This happens to the $W^\pm$ and $Z^0$
bosons: electroweak gauge symmetry is {\em spontaneously} broken in the
ground state of the theory, a phenomenon known as the ``Higgs
mechanism''.~\cite{higgs} Finally, fermions in the standard model also must
start out massless. To make quarks and leptons massive, new forces beyond
the $SU(3) \otimes SU(2) \otimes U(1)$ gauge interactions are required.
These additional interactions {\em explicitly} break the fermions' flavor
symmetry and communicate electroweak symmetry breaking to them.

Despite this great body of knowledge, the interactions underlying
electroweak and flavor symmetry breakdowns remain {\underbar{{\em
unknown}}. The most important element still missing from this description
of particle interactions is directly connected to electroweak symmetry
breaking. This may manifest itself as one or more {\em elementary} scalar
``Higgs bosons''. This happens in supersymmetry, the scenario for the
physics of electroweak symmetry breaking that is by far the most
popular.~\cite{time} Notwithstanding its popularity, there is no
experimental evidence for
supersymmetry.~\cite{searches}$^{,\ts}$~\footnote{Those who would cite the
apparent unification of the $SU(3) \otimes SU(2) \otimes U(1)$ couplings
near $10^{16}\,\gev$ as evidence now have to incorporate the scenario of
supersymmetry breaking mediated by new gauge interactions.} {\sl We do not
know the origin of electroweak symmetry breaking.}

If the dynamics of the Higgs mechanism are unknown in detail, those of
flavor are completely obscure. We don't even have a proper name, much less
a believable and venerable ``mechanism'', for flavor symmetry breaking.
Models with elementary Higgs bosons, whether supersymmetric or not, offer
no explanation at all for the quark-lepton content of the generations, the
number of generations, why they are identical, and why flavor symmetry is
broken---the bizarre pattern of quark and lepton masses.

Dynamical electroweak and flavor symmetry breaking---technicolor and
extended technicolor---are plausible, attractive, natural, and nontrivial
scenarios for this physics that involve new interactions at specified,
experimentally accessible energy scales.~\cite{tasi} Technicolor is a
strong gauge interaction modeled after QCD. Its characteristic energy is
$\simle 1\,\tev$, so it may be sought in experiments of the coming decade.
Extended technicolor (ETC) embeds technicolor, color and flavor into a
larger gauge symmetry; this embedding is necessary to produce the nonzero
``current-algebraic'' or ``hard'' masses of quarks and leptons. At the same
time, ETC offers a simple group-theoretic explanation of flavor in terms of
the representation content of fermions. As we explain shortly, the scale at
which ETC symmetry is broken down to color $\otimes$ technicolor is
$\CO(100\,\tev)$. Nevertheless, the effects of this interaction are
observable at the TeV~energy scale in terms of the masses and decay modes
of the technihadrons, $\rho_T$ and $\tpi$, that populate technicolor
models.

Because we are so completely ignorant of electroweak and flavor dynamics,
experiments at TeV~energies, which for now means those planned for the
Tevatron and the LHC, must have the greatest possible discovery potential.
They ought to search for technicolor and extended technicolor as well as
the standard model Higgs boson, its simple extensions, supersymmetry, and
so on. Hadron colliders have powerful reach by virtue of their high energy
and luminosity, but extracting clear signals from them can be quite
demanding. Thus, detectors should be designed to be sensitive to, and
experimenters should be prepared to search for, the signatures of dynamical
electroweak and flavor symmetry breaking. So far, there is little indication
of this in the large LHC detector collaborations.

\section{Summary of Technicolor and Extended Technicolor}\label{sec:techni}

Technicolor---the strong interaction of fermions and gauge bosons at the
scale $\LTC \sim 1\,\tev$---describes the breakdown of electroweak symmetry
to electromagnetism {\em without} elementary scalar bosons.~\cite{tcref}
Technicolor has a great precedent in QCD. The chiral symmetry of
massless quarks is spontaneously broken by strong QCD interactions,
resulting in the appearance of massless Goldstone bosons, $\pi$, $K$,
$\eta$.~\footnote{The hard masses of quarks explicitly break chiral
symmetry and give mass to $\pi$, $K$, $\eta$, which are then referred to as
pseudo-Goldstone bosons.} In fact, if there were no Higgs bosons, this
chiral symmetry breaking would itself cause the breakdown of
electroweak $SU(2) \otimes U(1)$ to electromagnetism. Furthermore, the $W$
and $Z$ masses would be given by $M_W^2 = \cos^2\theta_W M_Z^2 = \eighth
g^2 N_F f_\pi^2$, where $g$ is the weak $SU(2)$ coupling, $N_F$ the number
of massless quark flavors, and $f_\pi$, the pion decay constant, is only
$93\,\mev$.

In its simplest form, technicolor is a scaled up version of QCD, with
massless technifermions whose chiral symmetry is spontaneously broken at
$\LTC$. If left and right-handed technifermions are assigned to weak
$SU(2)$ doublets and singlets, respectively, then $M_W = \cos\theta_W M_Z =
\half g \Few$, where $\Few = 246\,\gev$ is the weak {\em technipion} decay
constant.~\footnote{The technipions in minimal technicolor are the linear
combinations of massless Goldstone bosons that become, via the Higgs
mechanism, the longitudinal components $W_L^\pm$ and $Z_L^0$ of the weak
gauge bosons.}

The principal signals in hadron collider experiments of ``classical''
technicolor were discussed long ago.~\cite{ehlq}$^{,\ts}$\cite{techniehlq}
In the minimal technicolor model, with just one technifermion doublet, the
only prominent collider signals are the modest enhancements in
longitudinally-polarized weak boson production. These are the $s$-channel
color-singlet technirho resonances near 1.5--2~TeV: $\troz \ra W_L^+W_L^-$
and $\tropm \ra W_L^\pm Z_L^0$. The $\CO(\alpha^2)$ cross sections of these
processes are quite small at such masses. This and the difficulty of
reconstructing weak-boson pairs with reasonable efficiency make observing
these enhancements a challenge.

Nonminimal technicolor models are much more accessible because they have a
rich spectrum of lower mass technirho vector mesons and technipion states
into which they may decay.~\footnote{The technipions of non-minimal
technicolor include the longitudinal weak bosons as well as additional
Goldstone bosons associated with spontaneous technifermion chiral symmetry
breaking. The latter must and do acquire mass---from the extended
technicolor interactions discussed below.} The often-discussed one-family
model, contains one isodoublet each of color-triplet techniquarks $(U,D)$
and color-singlet technileptons $(N,E)$. Because the color coupling is weak
above 100~GeV, the technifermion chiral symmetry is approximately $SU(8)
\otimes SU(8)$. This symmetry and its breakdown to the diagonal $SU(8)$
gives rise to 63~$\rho_T$ and $\tpi$ which may be classified according to
how they transform under ordinary color $SU(3)$ times weak isospin $SU(2)$.
The technipions are color singlets $\tpipr \in (1,1)$; $W^\pm_L, Z^0_L$ and
$\tpipm, \tpiz \in (1,3)$;  color octets $\etat \in (8,1)$ and $\octpipm,
\octpiz \in (8,3)$; and color-triplet leptoquarks $\tpiql,\ts \tpilq \in
(3,3) \oplus (3,1) \oplus (\ol 3,3)\oplus(\ol 3,1)$. The $\rho_T$ belong to
the same representations.

In the standard model and its extensions, the masses of quarks and leptons
are produced by their Yukawa couplings to the Higgs bosons---couplings of
arbitrary magnitude and phase that are put in by hand. This option is not
available in technicolor because there are no elementary scalars. Instead,
quark and lepton chiral symmetries must be broken explicitly {\it by gauge
interactions alone}. The most economical way to do this is to employ
extended technicolor, a gauge group containing flavor, color and
technicolor as subgroups. Quarks, leptons and technifermions are combined
into the same few large representations of ETC. Then quark and lepton hard
masses are generated by their coupling (with strength $\getc$) to
technifermions via ETC gauge bosons of generic mass $\METC$:
\be\label{qmass}
m_q(\METC) \simeq m_\ell(\METC)  \simeq {\getc^2 \over
{\METC^2}} \condetc \ts,
\ee
where $\condetc$ and $m_{q,\ell}(\METC)$ are the technifermion condensate
and quark and lepton masses renormalized at the scale $\METC$.

If technicolor is like QCD, with a running coupling $\atc$ rapidly becoming
small above $\LTC \sim 1\,\tev$, then $\condetc \simeq \condtc \simeq
\LTC^3$. To obtain quark masses of a few~GeV thus requires $\METC/\getc
\simle 30\,\tev$. This is excluded. Extended technicolor boson exchanges
also generate four-quark interactions which, generically, include $|\Delta
S| = 2$ and $|\Delta B| = 2$ operators. For these not to be in conflict
with $K^0$-$\ol K^0$ and $B_d^0$-$\ol B_d^0$ mixing parameters,
$\METC/\getc$ must exceed several hundred TeV.~\cite{etceekl} This implies
quark and lepton masses no larger than a few MeV, and technipion masses no
more than a few~GeV.

Because of this conflict between constraints on flavor-changing neutral
currents and the magnitude of ETC-generated quark, lepton and technipion
masses, classical technicolor was superseded a decade ago by ``walking''
technicolor.~\cite{wtc} Here, the strong technicolor
coupling $\atc$ runs very slowly---walks---for a large range of momenta,
possibly all the way up to the ETC scale of several hundred TeV. The
slowly-running coupling enhances $\condetc/\condtc$ by almost a factor of
$\METC/\LTC$. This, in turn, allows quark and lepton masses as large as a
few~GeV and $M_{\tpi} \simge 100\,\gev$ to be generated from ETC
interactions at $\METC = \CO(100\,\tev)$.

Walking technicolor requires a large number of technifermions in order that
$\atc$ runs slowly. These fermions may belong to many copies of the
fundamental representation of the technicolor gauge group, to a few higher
dimensional representations, or to both. That last possibility inspired
``multiscale technicolor'' models containing both fundamental and higher
representations, and having a very different phenomenology.~\cite{multi} In
multiscale models, there typically are two widely separated scales of
electroweak symmetry breaking, with the upper scale set by the weak decay
constant, $\Few = 246\,\gev$. Technihadrons associated with the lower scale
may be so light that they are within reach of the Tevatron collider; they
are readily produced {\em and detected} at the LHC.

An important consequence of walking technicolor is that the large ratio
$\condetc/\condtc$ significantly enhances technipion masses. Thus, $\rho_T
\ra \tpi\tpi$ decay channels may be closed. If this happens, then $\tro \ra
W_L W_L$ or $W_L \tpi$. The production rates for these color singlets are
5--10~pb at the Tevatron and 25--100~pb at the LHC. If colored
technifermions exist, the electrically neutral color-octet technirho,
$\troct$, may have its $\tpi\tpi$ decay channels closed as well. In this
case, it appears as a relatively narrow resonance in $\troct \ra$~dijets.
If the $\tro$, $\troct \ra \tpi\tpi$ channels are open, they are resonantly
produced at large rates, of order 5~pb at the Tevatron and several
nanobarns at the LHC. As we describe in more detail below, technipions tend
to decay to heavy fermions. Given these large rates and the recent
successes and coming advances in heavy flavor detection, many of these
technipions should be reconstructable in the hadron collider environment.

Another major development in technicolor was motivated by the recent
discovery of the top quark.~\cite{toprefs} Theorists have concluded that
ETC  models cannot explain the top quark's large mass without running afoul
of either experimental constraints from the $\rho$ parameter and the $Z \ra
\ol b b$ decay rate~\cite{blondel}$^{,\ts}$\cite{zbbth} (the ETC mass must
be about 1~TeV; see Eq.~(1)) or of cherished notions of
naturalness ($\METC$ may be higher, but the coupling $\getc$ must be
fine-tuned near to a critical value). This state of affairs has led to the
proposal of ``topcolor-assisted technicolor'' (TC2).~\cite{tctwohill}

In TC2, as in top-condensate models of electroweak symmetry
breaking,~\cite{topcondref} almost all of the top quark mass arises from a
new strong ``topcolor'' interaction.~\cite{topcref} To maintain electroweak
symmetry between (left-handed) top and bottom quarks and yet not generate
$m_b \simeq m_t$, the topcolor gauge group under which $(t,b)$ transform is
usually taken to be $SU(3)\otimes U(1)$. The $U(1)$ provides the difference
that causes only top quarks to condense. Then, in order that topcolor
interactions be natural---i.e., that their energy scale not be far above
$m_t$---without introducing large weak isospin violation, it is necessary
that electroweak symmetry breaking remain due mostly to technicolor
interactions.~\cite{tctwohill}

In TC2 models, ETC interactions are still needed to generate the light and
bottom quark masses, contribute a few~GeV to $m_t$,~\footnote{Massless
Goldstone ``top-pions'' arise from top-quark condensation. This ETC
contribution to $m_t$ is needed to give them a mass in the range of
150--250~GeV.} and give mass to the technipions. The scale of ETC
interactions still must be hundreds of~TeV to suppress flavor-changing
neutral currents and, so, the technicolor coupling still must walk. Early
steps in the development of the TC2 scenario have been taken in two recent
papers.~\cite{tctwoklee} Although the phenomenology of TC2 is in its
infancy, it is expected to share general features with multiscale
technicolor: many technihadron states, some carrying ordinary color, some
within range of the Tevatron, and almost all easily produced and detected
at the LHC at moderate luminosities.

I assume throughout this talk that the technicolor gauge group is $\sutc$
and that its gauge coupling walks. A minimal, one-doublet model can have a
walking $\atc$ only if the technifermions belong to a large non-fundamental
representation. For nonminimal models, I generally consider the
phenomenology of only the lighter technifermions. These transform according
to the fundamental~($\Ntc$) representation. Some of them may also be
ordinary color triplets. Finally, in TC2, there is no need for large
technifermion isospin splitting associated with the top-bottom mass
difference. This simplifies our discussion greatly.

The decays of technipions are induced mainly by ETC interactions which
couple them to quarks and leptons. These couplings are Higgs-like, and so
technipions are expected to decay into heavy fermion pairs. For the
color-singlets, e.g.,
\be\label{eq:singdecay}
\ba{ll}
\tpiz \ra &\left\{\ba{ll} b \ol b &\mbox{if $M_{\tpi} < 2 m_t$}  \\
t \ol t &\mbox{if $M_{\tpi} > 2 m_t$}
\ea \right.\\
\tpip \ra &\left\{\ba{ll} c \ol b \ts, \ts c \ol s, \ts \tau^+ \nu_\tau
&\mbox{if $M_{\tpi} < m_t + m_b$} \\
t \ol b &\mbox{if $M_{\tpi} > m_t + m_b$}
\ea \right.
\ea
\ee
An important exception to this rule occurs in TC2 models. There, only a
few~GeV of the top mass arises from ETC interactions. The $b \ol b$ mode of
a heavy $\tpiz$ then competes with $t \ol t$; $c \ol b$ or $c \ol s$
compete with $t \ol b$ for $\tpip$. Note that, since the decay $t \ra \tpip
b$ is strongly suppressed in TC2 models, the $\tpip$ can be much lighter
than the top quark.

In almost all respects, walking technicolor models are very different from
QCD with a few fundamental $SU(3)$ representations. One example of this is
that integrals of weak-current spectral functions and their moments
converge much more slowly than they do in QCD. Consequently, simple
dominance of spectral integrals by a few resonances cannot be correct. This
and other calculational tools based on naive scaling from QCD and on
large-$\Ntc$ arguments are suspect. Thus, it is not yet possible to predict
with confidence the influence of technicolor degrees of freedom on
precisely-measured electroweak quantities---the $S,T,U$ parameters to name
the most discussed example.~\cite{pettests}$^{,\ts}$~\footnote{These
comments respond to a question from Graham Ross regarding the effects of
technicolor on precision electroweak tests. I thank him for the opportunity
to reiterate them.~\cite{glasgow}}

\section{Technicolor Signatures at Hadron Colliders}\label{sec:signals}
\subsection{Color-Singlet Technipion Production}\label{sec:singlet}

The $\tro \ra W^+_L W^-_L$ and $W^\pm_L Z^0_L$ signatures of the minimal
technicolor model were discussed long
ago.~\cite{ehlq}$^{,\ts}$\cite{techniehlq} If there is to be just one
technifermion doublet, it must belong to a higher dimensional
representation of $\sutc$ so that $\atc$ walks. The main phenomenological
consequence of this is that it is questionable to use the $\tro \ra
\tpi\tpi$ coupling $\atro$ obtained by naive scaling from QCD,
\be\label{alpharho}
\atro = 2.91 \left({3\over{\Ntc}}\right)\ts.
\ee
This coupling may be smaller than Eq.~(3) indicates, leading
to a narrower $\tro$. There is also the possibility that, because of its
large mass (naively, 1.5--2~TeV), the $\tro$ has a sizable branching ratio
to four-weak-boson final states. To my knowledge, neither of these
possibilities has been investigated.

From now on, I consider only nonminimal models which, I believe, are much
more likely to lead to a satisfactory walking model. They have a rich
phenomenology with many diverse, relatively accessible signals. The masses
of technipions in these models arise from broken ETC and ordinary color
interactions. In walking models that have been studied,~\cite{multi} they
lie in the range 100--600~GeV; technirho vector meson masses are expected
to lie between 200 and 1000~GeV. Multiscale and topcolor-assisted models of
technicolor tend to have so many technifermions that the characteristic
scale of these models, set by the technipion decay constant $F_T$, is
small~\cite{multi}$^{,\ts}$\cite{tctwoklee}. Consequently, it is plausible
that technihadrons $\tpi$ and $\rho_T$ have masses at the lower end of
these ranges. I should not have to point out that such low-scale
technihadrons are accessible at the Tevatron.

Color-singlet technipions, including the longitudinal $W_L$ and $Z_L$, are
pair-produced via the Drell-Yan process in hadron collisions. The
$\CO(\alpha^2)$ signal rates at the Tevatron and LHC are probably
unobservably small compared to backgrounds {\em unless} there are fairly
strong color-singlet technirho resonances, $\tro^{\pm,0}$ not far above
threshold. To parameterize the cross sections, we consider a simple model
containing two isotriplets of technipions which are mixtures of $W_L^\pm$,
$Z_L^0$ and an isotriplet of mass-eigenstate technipions
$\tpi$.~\cite{multi}$^{,\ts}$\cite{tpitev} The lighter isotriplet $\tro$ is
assumed to decay dominantly into pairs of the mixed state
$\vert\Pi_T\rangle = \sin\chi \ts \vert W_L\rangle + \cos\chi \ts
\vert\tpi\rangle$, leading to the processes
\be\label{eq:singlet}
\ba{llll}
q \ol q'  \ra & W^\pm  \ra & \tropm \ra 
&\left\{\ba{l} W_L^\pm Z_L^0 \\ W_L^\pm \tpiz, \ts\ts \tpipm Z_L^0 \\
\tpipm \tpiz \ea \right. \\ \\
q \ol q   \ra & \gamma, Z^0  \ra & \troz \ra
&\left\{\ba{l} W_L^+ W_L^- \\ W_L^\pm \tpimp \\
\tpip \tpim \ea \right.
\ea
\ee

The mixing angle $\chi$ is specified by $\sin\chi = F_T/\Few$, where $F_T$
is the $\Pi_T$ decay constant and $\Few = 246\,\gev$. Although this mixing
usually is quite small, walking technicolor enhancements of technipion
masses suppress or even close the $\tro \ra \tpi\tpi$ channels. Thus, the
$\tro$ should be quite narrow and any of the decay modes in
Eq.~\ref{eq:singlet} may be important. This is seen in Fig.~1 where the
production rates of individual channels are calculated for the Tevatron as
a function of $M_{\tro}$ for $M_{\tpi} = 110\,\gev$ and $\sin\chi =
\third$. Such low mass technipions are expected to decay to $b \ol b$ or $c
\ol b$. Furthermore, some scheme such as TC2 which results in a small
coupling $m^{ETC}_t/F_T$ of technipions to the top quark is required to
suppress the unseen mode $t \ra \tpip b$.~\cite{twbrate} Heavy-flavor
tagging and kinematical selection techniques are useful to extract the
signals. Figure~1 illustrates several important general
points:~\cite{tpitev}$^{,\ts}$\cite{elw}

\vfil\eject

\begin{figure}
\vskip -1.3truein
%\center
%\rule{2cm}{0.2mm}\hfill \rule{2cm}{0.2mm}
%\vskip 2.8cm
%\rule{2cm}{0.2mm}\hfill \rule{2cm}{0.2mm}
\psfig{figure=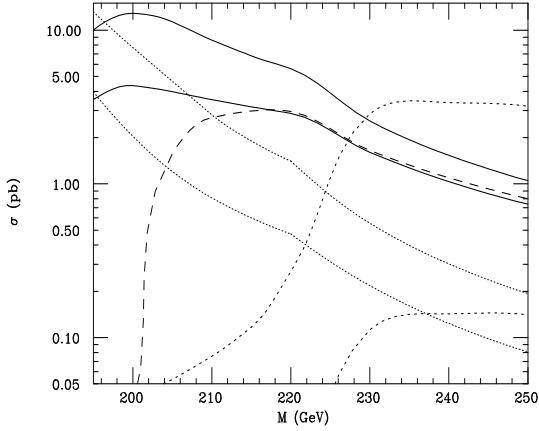,height=3.8in,width=2.5in}
\caption{Total $WW$, $W\tpi$ and $\tpi\tpi$ cross sections in $\ol p p$
collisions at $1.8\,\tev$, as a function of $M_{\tro}$ for $M{\tpi} =
110\,\gev$. The model described in the text with $\sin\chi = \third$ is
used. The curves are $W^\pm Z^0$ (upper dotted) and $W^+W^-$ (lower
dotted); $W^\pm \tpiz$ (upper solid), $W^\pm \tpi^\mp$ (lower solid),
and $Z^0\tpi^\pm$ (long dashed); $\tpi^\pm \tpiz$ (upper short dashed)
and $\tpip\tpim$ (lower short dashed).}
\label{fig:tpitevfig}
\end{figure}

\begin{itemize}

\item Except near $W\pi_T$ threshold, the increase in $WZ$ and $WW$
production is undetectably small.

\item The most important processes are those with positive $Q = [M_{\tro}$
-- (sum of final state masses)$]$ {\it and} the fewest number of
longitudinal weak bosons. At the Tevatron, the inclusive $W\tpi$ rate is
5--$10\,\pb$ and the $Z\tpi$ rate is 1--$3\,\pb$ for $M_{\tpi} + M_W \simle
M_{\tro} \simle 2M_{\tpi}$. These rates are 5--10 times larger at the LHC.
Because the $\tro$ is very narrow, the $\tpi \ra$~dijet system should have
large azimuthal opening angle $\Delta\phi(jj) \simge 125^\circ$ and limited
transverse momentum $p_T(jj) \simle (M^4_{\tro} - 2(M_{\tpi} + M_W)^2
M^2_{\tro} + (M^2_{\tpi} - M^2_W)^2)^{\half} / 2M_{\tro} \simeq 50\,\gev$.

\item Signal events for $W/Z + \tpi$ should exhibit a narrow peak,
consistent with resolution, corresponding to the $\tro$ resonance. This,
however, may not be a good way to discriminate signal from background
because kinematic cuts can ``sculpt'' such a peak.

\item Once $M_{\tro} \ge 2 M_{\tpi} + 10\,\gev$, the dominant process is
$\tpipm\tpiz$ production. The crossover point depends to some extent on the
suppression factor $\tan\chi$, but it should not be much different from
this. A search for the $\tpipm\tpiz$ channel will be rewarding, even if it
is negative.

\end{itemize}

Since the isospin of technifermions is approximately conserved, the $\tro$
is expected to be nearly degenerate with its isoscalar partner $\omega_T$.
The walking technicolor enhancement of technipion masses almost certainly
closes off the isospin-conserving decay $\omega_T \ra \Pi^+_T \Pi^-_T
\Pi^0_T$. Even the triply-suppressed mode $W^+_L W^-_L Z_L$ has little or
no phase space for $M_{\omega_T} \simle 300\,\gev$. Thus, the main decays
are expected to be $\omega_T \ra \gamma \Pi_T^0$, $Z \Pi_T^0$, and $\Pi_T^+
\Pi_T^-$. In terms of  mass eigenstates, these modes are $\omega_T \ra
\gamma \tpiz$, $\gamma Z_L$, $Z \tpiz$, $Z Z_L$; $\gamma \tpipr$, $Z
\tpipr$; and $W^+_L W^-_L$, $\tpipm W^\mp_L$, $\tpip\tpim$.~\footnote{The
modes $\omega_T \ra \gamma Z_L$, $Z Z_L$ were considered by Chivukula and
Golden for a one-doublet technicolor model.~\cite{cg} Our estimates of the
branching ratios for the isospin-violating decays $\tro \ra \gamma \tpiz$,
$Z \tpiz$ suggest that they are negligible unless the mixing angle $\chi$
is very small.} It is not possible to estimate the relative magnitudes of
the decay amplitudes without an explicit model of the $\omega_T$'s
constituent technifermions. Judging from the decays of the ordinary
$\omega$, we expect $\omega_T \ra Z \tpiz (\tpipr)$, $\gamma \tpiz
(\tpipr)$ to dominate, with the latter mode favored by phase space.

The $\omega_T$ is produced in hadron collisions just as the $\tro^0$ is,
via its vector-meson-dominance coupling to $\gamma$ and $Z^0$. For
$M_{\omega_T} \simeq M_{\tro}$, the $\omega_T$ production cross section
should be approximately $|Q_U + Q_D|^2$ times the $\tro^0$ rate, where
$Q_{U,D}$ are the electric charges of the $\omega_T$'s constituent
technifermions. The principal signatures for $\omega_T$ production, then,
are $\ell^+\ell^-$ (or $\nu \ol \nu$) $+ b \ol b$ and $\gamma + \ol b b$,
with $M_{\ol b b} = M_{\tpi}$. A search for the $Z\tpi$ mode will use the
same strategies as for $\tro \ra Z_L \tpi$ and $W_L \tpi$. The search for
$\omega_T \ra \gamma \tpi$ in hadron collider experiments is under
study.~\cite{elw}

\subsection{Color-Octet Technirho Production and Decay to Jets and
Technipions}\label{sec:octet} 

Models with an electroweak doublet of color-triplet techniquarks $(U,D)$
have an octet of $I=0$ technirhos, $\troct$, with the same quantum numbers
as the gluon. The $\troct$ is produced strongly in $\ol q q$ and $gg$
collisions. Assuming the one-family model for simplicity, the 63
technipions listed in Sec.~\ref{sec:techni} also occur. There are two
possibilities for $\troct$ decays.~\cite{multi}

In the first, walking technicolor enhancements of the technipion masses
close off the $\tpi\tpi$ channels. Then the octet technirho's coupling to
the gluon mediates $\troct \ra \ol q q,\ts gg \ra \jets$. The
$\CO(\aqcd^2)$ dijet cross sections including the $\troct$ enhancement are
illustrated for the Tevatron and the LHC in Figs.~2
and~3.~\cite{multi}$^{,\ts}$\cite{snow} For $M_{\troct} = 250\,\gev$, the
signal-to-background rates is estimated to be 0.70~nb/5.0~nb at the
Tevatron and 15~nb/150~nb at the LHC. For $M_{\troct} = 500\,\gev$, the
$S/B$ rates in these figures are 10~pb/40~pb and 2.0~nb/6.0~nb,
respectively. Searches for the dijet signal of $\troct$ have been carried
out by the CDF Collaboration.~\cite{cdfdijet}$^{,\ts}$\cite{rharris_jet}
Using $103\,\ipb$ of data from Tevatron Collider Run~I, CDF has excluded
the range $250\,\gev < M_{\troct} < 500\,\gev$ for the model~A parameters
used in the second paper of Ref.~15.

\begin{figure}
\vskip -1.3truein
%\center
%\rule{2cm}{0.2mm}\hfill \rule{2cm}{0.2mm}
%\vskip 2.8cm % \rule{2cm}{0.2mm}\hfill \rule{2cm}{0.2mm}
\psfig{figure=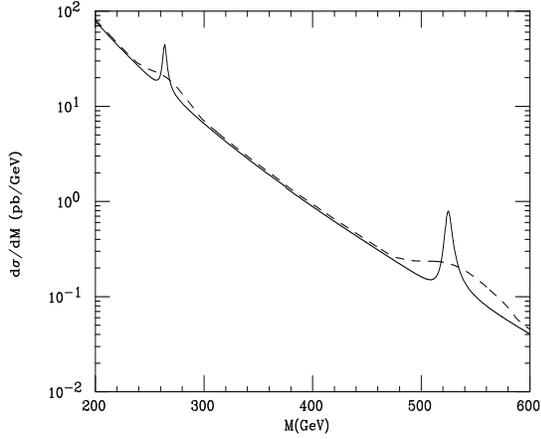,height=3.8in,width=2.5in}
\caption{Dijet cross sections at the Tevatron ($\ol p p$ collisions at
1.8~TeV) including the effect of octet techirho vector mesons at 250 and
500~GeV. The solid curve assumes perfect jet energy resolution while the
dashed curve assumes resolution $\sigma(E)/E = 100\% E^\half(\gev)$. Jet
angles and rapidities were limited by $\cos \theta^* < \twothirds$ and
$\vert \eta_j \vert < 2.0$.}
\label{fig:tevjetfig}
\end{figure}

\begin{figure}
\vskip -1.3truein
%\center
% \rule{2cm}{0.2mm}\hfill \rule{2cm}{0.2mm}
% \vskip 2.8cm % \rule{2cm}{0.2mm}\hfill \rule{2cm}{0.2mm}
\psfig{figure=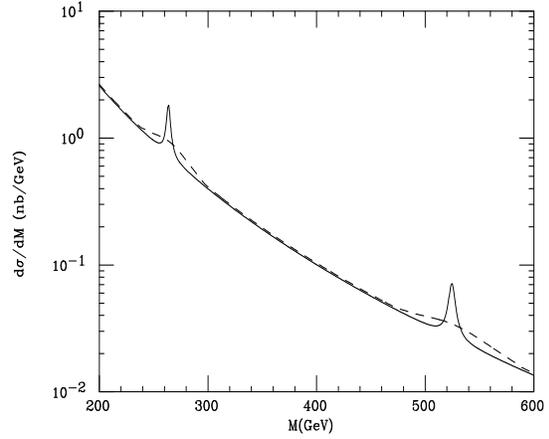,height=3.8in,width=2.5in}
\caption{Dijet cross sections at the LHC ($p p$ collisions at
14~TeV) including the effect of octet techirho vector mesons at 250 and
500~GeV. Resolutions and cuts are as in Fig.~2 except that $\vert \eta_j
\vert < 1.0$.}
\label{fig:lhcjetfig}
\end{figure}

The second possibility is that technipion decay channels are open, in which
case $\troct \ra \octpi \ol \octpi$ and $\tpiql\tpilq$ dominates the dijet
modes. The color octet technipions are expected to decay into heavy quark
pairs, as do the color-singlets in Eq.~\ref{eq:singdecay}. The
color-triplet leptoquarks decay as
\be\label{eq:lqdecay}
\ba{ll}
\tpiun \ra &\left\{\ba{ll} c \ol \nu_\tau  &\mbox{if $M_{\tpi} <  m_t$} \\
t \ol \nu_\tau &\mbox{if $M_{\tpi} >  m_t$}
\ea \right.\\
\tpiue \ra &\left\{\ba{ll} c \tau^+ &\mbox{if $M_{\tpi} < m_t$} \\
t \tau^+ &\mbox{if $M_{\tpi} > m_t$} \ea \right. \\
\tpidn \ra & b \ol \nu_\tau \\
\tpide \ra & b \tau^+ \ts. 
\ea
\ee
The caveat regarding technipion decays to top quarks in TC2 models still
applies. Technipion pair production rates, per channel, are expected to
lie in the range 1--10~pb at the Tevatron and 1--10~nb at the LHC.
Detailed rate estimates depend on $\troct$ and $\tpi$ masses and other
model parameters. The LHC rate estimates are so high that color octet and
triplet technipions cannot fail to be discovered there---if they exist and
if they have not already been detected in Run~II of the Tevatron.

At this conference, K.~Maeshima of CDF reported on a search for
color-triplet leptoquarks decaying into $\tau+\ts \jet$.~\cite{kaori} The
limit obtained, $M_{\tpiql} \simge 100\,\gev$ assumes only pure-QCD
production of the leptoquark pair. A somewhat more stringent (and more
model-dependent) limit would result if it is assumed that the leptoquarks
are resonantly produced.

\subsection{Signatures of Topcolor-Assisted Technicolor}\label{sec:tctc}

Topcolor and topcolor-assisted technicolor (TC2) were reviewed at this
conference by D.~Kominis.~\cite{kominis} The development of TC2 is still at
an early stage and, so, its phenomenology is not fully formed.
Nevertheless, in addition to the color-singlet and nonsinglet technihadrons
already discussed, there are three TC2 signatures that are likely to be
present in any surviving model:~\cite{tctwohill}$^{,\ts}$\cite{tctwoklee}
$^{,\ts}$\cite{hp}$^{,\ts}$\cite{snow}

\begin{itemize}

\item The isotriplet of color-singlet ``top-pions'' $\toppi$ arising
from spontaneous breakdown of the top quark's $SU(2)\otimes U(1)$ chiral
symmetry.

\item The color-octet of vector bosons $V_8$, called ``colorons'',
associated with breakdown of the top quark's strong $SU(3)$ interaction to
ordinary color.

\item The $Z'$ vector boson associated with breakdown of the top
quark's strong $U(1)$ interaction to ordinary weak hypercharge.

\end{itemize}

\medskip

The three top-pions are nearly degenerate. They couple to the top quark
with strength $m_t/F_t$, where $m_t$ is the part of the top-quark mass
induced by topcolor---expected to be within a few GeV of its total
mass---and $F_t \simeq 70\,\gev$~\cite{tctwohill} is the $\toppi$ decay
constant. If the top-pion is lighter than the top quark, then
\be\label{eq:tpibrate}
\Gamma(t \ra \toppip b) \simeq
{(m_t^2 - M_{\pi_t}^2)^2 \over {16 \pi m_t F_t^2}} \ts.
\ee
The standard top-decay mode branching ratio $B(t \ra W^+ b) =
0.87^{+0.13}_{-0.30}$ (stat.)~$^{+0.13}_{-0.11}$ (syst.)~was reported
a year ago.~\cite{twbrate} At the $1\sigma$ level, then, $M_{\pi_t} \simge
150\,\gev$. At the $2\sigma$ level, the lower bound is $100\,\gev$, but
such a small branching ratio for $t \ra W^+ b$ would require $\sigma(p\ol p
\ra t \ol t)$ at the Tevatron about~4 times the standard QCD value of
$4.75^{+0.63}_{-0.68}\,\pb$.~\cite{topsig} The $t \ra \toppip b$ decay mode
can be sought in high-luminosity runs at the Tevatron and with moderate
luminosity at the LHC. If $M_{\pi_t} < m_t$, then $\toppip \ra c \ol b$
through $t$--$c$ mixing. It is also possible, though unlikely, that $\tpip
\ra t \ol s$ through $b$--$s$ mixing.

If $M_{\pi_t} > m_t$, then $\toppip \ra t \ol b$ and $\toppiz \ra \ol t t$
or $\ol c c$, depending on whether the neutral top-pion is heavier or
lighter than $2m_t$. The main hope for discovering top-pions heavier than
the top quark seems to rest on the isotriplet of top-rho vector mesons,
$\rho_t^{\pm,0}$. It is hard to estimate $M_{\rho_t}$; it may lie near
$2m_t$ or closer to $\Lambda_t = \CO(1\,\tev)$. The $\rho_t$ are produced
in hadron collisions just as the corresponding color-singlet technirhos
discussed above. The conventional expectation is that they decay as
$\rho_t^{\pm,0} \ra \toppipm \toppiz$, $\toppip \toppim$. The rates are not
likely to be large, but the distinctive decays of top-pions help suppress
standard model backgrounds.

It is also plausible that, because topcolor is broken near $\Lambda_t$, the
$\rho_t$ are not completely analogous to the $\rho$-mesons of QCD and
technicolor. For distance scales between $\Lambda^{-1}_t$ and
$1\,\gev^{-1}$, top and bottom quarks do not experience a growing confining
force. Instead of $\rho_t \ra \toppi\toppi$, the $\rho_t^{\pm,0}$ may fall
apart into their constituents $t \ol b$, $b \ol t$ and $t \ol t$. The
$\rho_t^\pm$ resonance may be visible as a significant increase in $t \ol
b$ production, but $\rho_t^0$ won't be seen in $t \ol t$.

The $V_8$ colorons of broken $SU(3)$ topcolor are readily produced in
hadron collisions. They are expected to have a mass of 0.5--1~TeV.
Colorons couple with strength $-g_S \cot\xi$ to quarks of the two light
generations and with strength $g_S \tan\xi$ to top and bottom quarks, where
$\tan\xi \gg 1$.~\cite{hp} Their decay rate is
\be\label{eq:widveight}
\ba{ll}
\Gamma_{V_8} = &\sixth \aqcd \Mv \ts
\biggl\{4\cot^2\xi\\
&+ \tan^2\xi \left(1 + \beta_t
(1-m_t^2/\Mv^2)\right)\biggr\} \ts,
\ea
\ee
where $\beta_t = \sqrt{1 - 4 m_t^2/\Mv^2}$. Colorons may then appear as
resonances in $b \ol b$ and $t \ol t$ production. R.~Harris has studied the
limits on masses and couplings of colorons decaying to $\ol b b$ and $\ol t
t$ that may be set at the Tevatron in Run~II ($\int \CL dt = 2\,\fb^{-1}$)
and at the high-luminosity TeV33 upgrade ($\int \CL dt = 30\,\fb^{-1}$). He
found that nearly the entire interesting range, $\Mv \simle 1.3\,\tev$, can
be probed.~\cite{rmhcoloron}

Colorons have little effect on the standard dijet production rate. The
situation may be very different for the $Z'$ boson of the broken strong
$U(1)$ interaction. In some TC2 models~\cite{tctwoklee} it is natural that
$Z'$ couples strongly to the fermions of the first two generations as well
as those of the third. The $Z'$ in these models is heavier than the
colorons, roughly $\Mzp =1$--$3\,\tev$.~\cite{rscjt} Thus, at subprocess
energies well below $\Mzp$, the interaction of $Z'$ with all quarks is
described by a contact interaction, just like what is expected for quarks
with substructure at a scale of a few~TeV. This leads to an excess of jets
at high $E_T$ and invariant mass.~\cite{elp}$^{,\ts}$\cite{ehlq} An excess
in the jet-$E_T$ spectrum consistent with $\Lambda \simeq 1600\,\gev$ has
been reported by the CDF Collaboration.~\cite{cdfjets}$^{,\ts}$\cite{brock}
It remains to be seen whether it is due to topcolor or any other new
physics. As with quark substructure, the angular and rapidity distributions
of the high-$E_T$ jets induced by $Z'$ should be more central than
predicted by QCD. The $Z'$ may also produce an excess of high invariant
mass $\ell^+\ell^-$. It will be interesting to compare limits on contact
interactions in the Drell-Yan process with those obtained from jet
production.

If the $Z'$ is strongly coupled to light fermions it will be produced
directly in $\ol q q$ annhilation in LHC experiments.  Because it may be
strongly coupled to so many fermions, including technifermions in the LHC's
energy range, it is likely to be very broad. This possibility should be
taken into account in forming strategies to look for the topcolor $Z'$ at
the LHC. I reiterate, however, that it is too early to predict the $Z'$
couplings, width and branching fractions with confidence. I hope for
progress on these questions in the coming year.

\section{Conclusions}\label{sec:close}

In this talk, I have tried to emphasize the importance of searching for
signatures of dynamical as well as weakly-coupled scenarios for electroweak
and flavor physics. We cannot be reminded too often how unaware we remain
of TeV-scale physics and that only experiment will remove our ignorance. A
young woman in Amherst, Massachusetts, said it best over a century ago:

\medskip

\begin{verse}
``Faith'' is a fine invention \\
When Gentlemen can see ---

But  {\it Microscopes}  are prudent \\
In an Emergency.

\vskip0.15truein

\hskip0.5truein --- {\sl Emily Dickinson, 1860}

\end{verse}

\vfil

\section*{Acknowledgments}

I thank Barry Barish and Paul Frampton for thoughtful comments on my talk.
I also thank Elizabeth Simmons for her careful reading and comments on the
manuscript. I thank the organizers, especially Andrzej Wroblewski, for a
wonderful conference, the opportunity to visit Poland, and much patience
and help. I owe Chris Quigg much for the opportunity to share my views on
the search for TeV~scale physics. My research was supported in part by the
Department of Energy under Grant~No.~DE--FG02--91ER40676.

\vfil

\section*{Questions}
\noindent{\it David J.~Miller, University College, London:}

SUSY has the attraction that she [sic] offers us a dark matter candidate.
Does your technicolor theory?

\vskip 12pt
\noindent{\it K.~Lane:}

The topcolor-assisted technicolor models I have been investigating tend to
have stable technibaryons. Whether or not they are electrically charged
and, hence, ruled out is a model-dependent question.

\vskip 12pt
\noindent{\it Bernd Kniel, Max Planck Institute, Munich:}

Technifermions have gauge couplings and introduce thresholds in the beta
functions of the gauge couplings. In addition, there is a new gauge
coupling introduced by technicolor. Will these couplings meet at some grand
unification scale as they do in supersymmetry?

\vskip 12pt
\noindent{\it K.~Lane:}

There already is a ``petit unification''---at the extended technicolor
scale of several hundred TeV where technicolor, color, and flavor gauge
symmetries are rejoined. I have no idea whether technicolor will involve
grand unification at some very high scale, although the attractiveness of
that possibility is undeniable. I remind you that the modern gauge-mediated
scenarios of supersymmetry breaking introduce new gauge couplings and, so,
also imperil the grand unification claimed for the $SU(3) \otimes SU(2)
\otimes U(1)$ couplings. In view of the recent developments in duality, I
would not be surprised in ten years time to find supersymmetry and
technicolor united into a strong-dynamical theory of electroweak and flavor
symmetry breaking, with quarks and leptons as composite entities at some
scale.

\end{document}